# Interleaving Commands: a Threat to the Interoperability of Smartcard Based Security Applications

Maurizio Talamo, Maulahikmah Galinium, Christian H. Schunck, Franco Arcieri

*Abstract*— Although smartcards are widely used, secure smartcard interoperability has remained a significant challenge. Usually each manufacturer provides a closed environment for their smartcard based applications including the microchip, associated firmware and application software. While the security of this "package" can be tested and certified for example based on the Common Criteria, the secure and convenient interoperability with other smartcards and smartcard applications is not guaranteed. Ideally one would have a middleware that can support various smartcards and smartcard applications. In our ongoing research we study this scenario with the goal to develop a way to certify secure smartcard interoperability in such an environment. Here we discuss and experimentally demonstrate one critical security problem: if several smartcards are connected via a middleware it is possible that a smartcard of type *S* receives commands that were supposed to be executed on a different smartcard of type *S'*. Such "external commands" can interleave with the commands that were supposed to be executed on *S*. Here we demonstrate this problem experimentally with a Common Criteria certified digital signature process on two commercially available smartcards. Importantly, in some of these cases the digital signature processes terminate without generating an error message or warning to the user.

*Keywords* — Common criteria, digital signature, interoperability, smartcard

## I. INTRODUCTION

Smartcards (SC) are becoming increasingly popular in many countries and are deployed, for example, as credit cards, health cards [1], public transportation service cards [2] and electronic identification documents. With these devices users control highly sensitive information and may perform security tasks such as mobile application security [3], electronic authentication and digital signature [4], [5]. As the importance and world-wide spread of SCs increases, the interoperability of these devices becomes more important along with their security in environments where SCs from different manufacturers and issuers are used at the same time.

The Common Criteria (CC) [6] and the CWA 14169 [7] standards are used to certify the correct behavior of a SC in a well defined environment, i.e. for a specific target of evaluation (TOE). The TOE is precisely described and usually comprises a specific microprocessor, a specified firmware and specified middleware [6]; however, environments where different SCs are used at the same time are usually different from the TOE. In particular, a SC could be confronted with commands from different processes, be it accidentally, on purpose or during an attack. Trusted SC interoperability, therefore, requires a careful analysis of how SCs operate in such situations and the consideration of these results in the design of interoperable systems.

One goal of current research and development efforts regarding SC interoperability is to create a framework that enables the concurrent use of different SCs. These efforts focus on diverse topics such as standardization [8], [9], [10], architectures for SC based authentication services [11], public key infrastructure [12], [13] and open protocols [14]. For example reference [15] studied the SC interoperability on public transit fare payment application using contactless SC. The authors propose a new payment protocol to support interoperability among different electronic purses and PSAMs (Purchase Secure Application Module) issued by different manufacturers.

When several SCs are connected to SC application via a middleware commands which are supposed to be executed on a certain SC may in fact be executed on a different one. Fig. 1 illustrates this situation: SC applications give input to and receive an output from SCs sharing a common middleware. The middleware translates the input into command sequences, i.e. into straight line program (SLP) which are supposed to be executed on a corresponding SC. The security problem mentioned above is indicated in the figure by the dashed arrows: commands interleave between the straight line programs. As a result a command may be executed on a SC different from the intended one. While such situations may arise inadvertently due to potential errors in the middleware such vulnerabilities can also be exploited in an attack. From a security perspective such events are particularly problematic if the SC executing a misdirected command does not immediately return an error message. We have observed this problem experimentally [4] and refer to such a situation as an ``anomaly''.

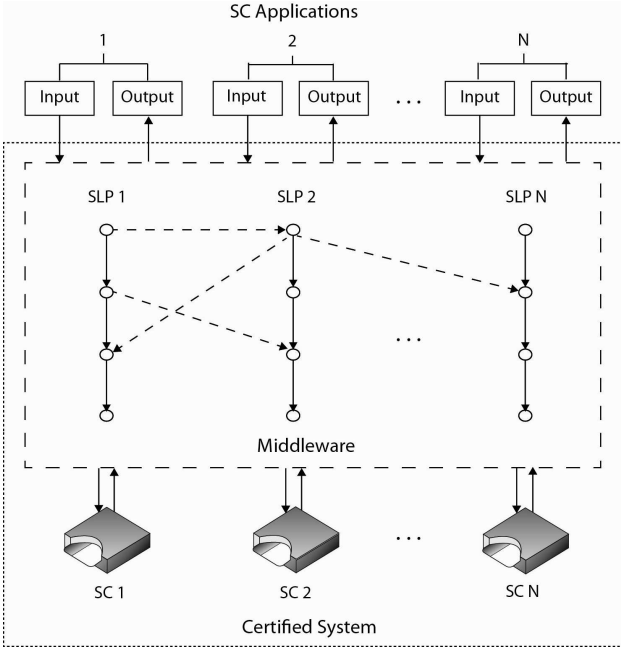

Fig. 1 Middleware

However, interoperability problems emerge already with the attempt to recognize what type of SC is actually used. In practice, this is currently done by detecting the presence of certain applications on the SC [16]. Deducing the SC type from such information is, at its best, an indirect method which might not uniquely identify the SC type, and leave it prone to potential attacks. If the SC type is incorrectly identified, "external commands" will be sent to the SC. Here, we define an "external command" as an Application Protocol Data Unit (APDU) sequence that does not correspond to the regular APDU sequence supplied to the SC in the executable code of the middleware originally used during the security certification (e.g. according to CC). In a setting where different SCs interact with applications via a middleware, APDUs that are supposed to be delivered to a certain SC type $S$ might be received by a SC of type $S'$ (e.g., due to routing errors). In such situations "external commands" can interleave with regular commands.

References [17], [18] study the behavior of commercial signature SCs during the sequential steps of a digital signature process. First the APDUs sent - in the setting used for the CC certification - from the middleware to the SC were identified. Using a model checking approach, the SCs were then targeted with modified APDUs during the digital signature process of a fixed document. The experiments showed that certain modified commands are accepted by the SCs without errors being generated and demonstrated that CC certification is not sufficient to address the SC interoperability problem.

In this paper we address the problem of interleaving commands for SC interoperability by analyzing the situation in which different applications interact with SCs via a middleware. A CC certified digital signature process on a commercially available SC is then tested to demonstrate the relevance of this problem experimentally. Finally, we discuss the complexity of the underlying issues and how the experimental test setup may be improved in the future to identify and prevent potential interoperability problems of this kind.

## II. THE INTEROPERABILITY PROBLEM: INTERLEAVING COMMAND SEQUENCES

To address the interoperability problem on a fundamental level, we consider a straight-line program $P_1$ with steps $S_{1,1}$, $S_{1,2}$, ..., $S_{1,l}$. It is assumed that the straight line program $P_1$ has been certified to produce a correct result if the sequence of commands $C_{1,1}$, $C_{1,2}$, ..., $C_{1,l}$ originally associated with these steps is executed in the correct order and without modifications on a SC of type $S$. For example, the command sequence $C_{1,1}$, $C_{1,2}$, ..., $C_{1,l}$ could match the one supplied to S in the executable code of the middleware that was used for CC certification. We call a command $C_{i,j}$ "globally legal" if it is processed in step $S_{i,j}$ of program $P_i$ on SC of type $S$ and the process $P_i$ has been certified on $S$.

In an environment where several applications interact with SCs via a middleware, commands from another straight line program $P_2$ may interleave with the commands from $P_1$ on $S$. Here $P_2$ is a straight line program for another SC type $S'$ with steps $S_{2,1}$, $S_{2,2}$, ..., $S_{2,k}$ and commands $C_{2,1}$, $C_{2,2}$, ..., $C_{2,k}$. Fig. 2 illustrates the situation and shows how SC $S$ receives interleaving commands associated with different steps of the two digital signature processes $P_1$ and $P_2$.

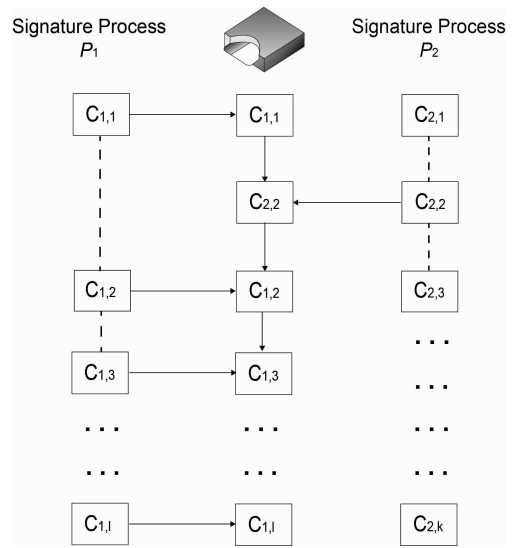

Fig. 2 Two Concurrent Signing Sessions

We will now analyze this SC interoperability problem in more detail, and in particular we distinguish the following cases:

1) In step $S_{1,j}$ of $P_1$, SC $S$ receives command $C_{1,j}$ and processes it without error. The digital signature process makes a correct transition to step $S_{1,j+1}$.

2) In step $S_{1,j}$ of $P_1$, SC $S$ receives command $C_{2,i}$ corresponding to step $S_{2,i}$ of $P_2$ and correctly generates an error. At this point the error can be detected and the digital signature process can be interrupted.

3) In step $S_{1,j}$ of $P_1$, SC $S$ receives command $C_{2,i}$ corresponding to step $S_{2,i}$ of $P_2$ and processes it without generating an error, i.e. the SC recognizes this command as *"locally legal"*. However, in this case, $C_{2,i}$ is *not* globally legal. We refer to this situation as an *"anomaly"* since it is unknown how the overall signature process will be affected. The program may now potentially make a transition to any step of the two programs $P_1$ or $P_2$.

Fig. 3 shows the first case that simply describes the correct process $P_1$ without interleaving commands from process $P_2$.

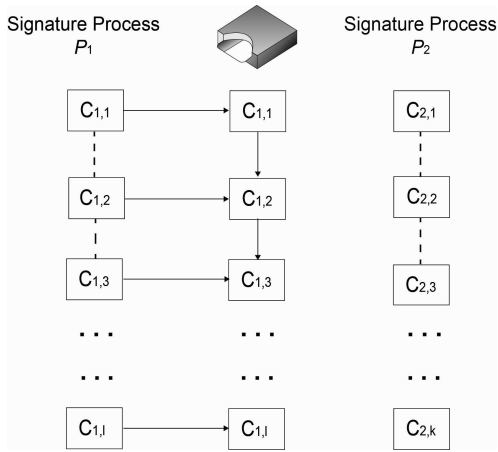

Fig. 3 Correct Process

Fig. 4 illustrates the second case: the digital signature process $P_1$ is interrupted by another process $P_2$, but an error is generated and this error can be detected by the middleware. The process $P_1$ can be terminated in this case. The interoperability environment can then be designed to handle such situations appropriately.

The third possibility (see fig. 2), however, poses the real problem for trusted SC interoperability: the "certified" and, therefore, trusted process has been modified but no error message has been generated. One anomaly can potentially be followed by several others and finally the digital signature process may terminate with a questionable result. Without receiving an error or a warning, a user cannot know whether all steps in the digital signature process were completed correctly or whether there have been one or more anomalies.

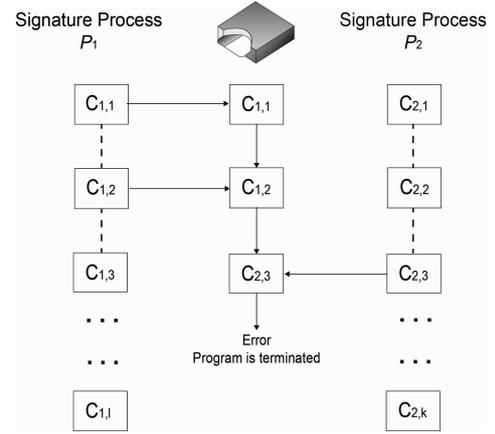

Fig. 4 An interleaving command results in an error

In the following we demonstrate experimentally, with a commercially available Common Criteria certified SC for digital signature, how commands from a different straight line program may interleave with the original one. Furthermore, we present one example where even though an error is generated an external command that intersects the original program can render a SC inappropriate for further use. The testing environment developed for this purpose, as well as relevant details about SCs are described in the next section.

### III. THE TESTING ENVIRONMENT

In our experiments, we study two commercially available SCs from two different manufacturers. The core of a SC is its microprocessor, which contains on board, a cryptographic processor, a small EEPROM random access memory ($\approx 64$ KBytes), an operating system and a memory mapped file system [19]. The microprocessor is customized (masked) in order to execute APDU sent from external software applications through a serial communication line.

The ISO 7816 standard [8], specifies the set of APDU that can be implemented by any compatible SC microprocessor. In particular, an APDU consists of a mandatory header of 4 bytes: the Class Byte (cla), the Instruction Byte (ins) and two parameter bytes (p1, p2). The header can be followed by a conditional body of variable length, which is composed by the length (in bytes) of the data field (lc), the data field itself and the maximum number of bytes expected in the data field of the response (le). Responses to any APDU are encoded in a variable length data field and two bytes mandatory return codes.

To probe and analyze the SC behavior we have developed a Crypto Probing System (CPS) whose overall architecture is shown in fig. 5. As each SC uses a different APDU sequence in the digital signature process, the CPS is designed to interface with both SCs used in this project. Effectively, it therefore acts as a *middleware* between the external applications and the real SCs.

The CPS is able to translate its simplified instructions to the corresponding sequence of APDUs (cla, ins, p1, p2, length and values of the possible annexed data buffer) to be sent to the connected physical SC and to translate the SC responses in a common format. Moreover, to further simplify the interface with the SC, the CPS is given the *globally legal* APDUs to be sent in each step of the digital signature process (SC commands flow), and the CPS is able to generate alternate command sequences to test the SC responses in different situations. This way, the CPS offers a simple interface for testing applications verifying process correctness and robustness on different physical devices and in the presence of interleaving command sequences.

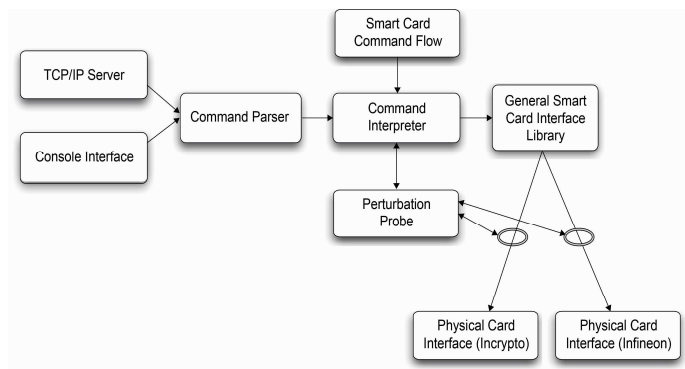

Fig. 5 Architecture of the Crypto Probing System

The CPS can be invoked via command line, to interactively test the command sequences, or used as a daemon, which stays in execution and accepts commands on TCP/IP connections. The commands sent via command line are parsed and interpreted by the CPS based on SC library. The elementary instruction of the CPS is made by a single APDU.

To meet the requirements of a complete digital signature process, *globally legal* APDUs for the Infineon-CardOS SC involve the following steps:
- step 0 Reset the SC
- step 1 Change directory to the Master File or root directory of the SC file system (SELECT FILE command)
- step 2 Activate the security environment for the digital signature (MSE-Manage Security Environment RESTORE command)
- step 3 At file system level, choose the private key to be used into the activated security environment (MSE SET command)
- step 4 Transmit the PIN connected to the private key used for the digital signature operation (VERIFY PIN command)
- step 5 Compute and send the data buffer ciphered using the selected private key and receive the result signature (PSO CDS - Perform Security Operation Compute Digital Signature command)

*Globally legal* APDUs for the Incrypto SC involve the following steps [5]:
- step 0 Reset the SC
- step 1 Change directory to the Master File or root directory of the SC file system (SELECT FILE command)
- step 2 Change to subdirectory containing the digital signature certificate will be used (SELECT FILE command)
- step 3 Activate the security environment for the digital signature (MSE-Manage Security Environment RESTORE command)
- step 4 At file system level, choose the private key to be used into the activated security environment (MSE SET command)
- step 5 Ask the SC for the random number that will be used as challenge for the next commands. It is first step for the activation of Secure Messaging (GET CHALLENGE command)
- step 6 Transmit a random number to the SC as a challenge for the next commands. It is the second step for the activation of Secure Messaging (GIVE CHALLENGE command)
- step 7 Transmit, using the two random number previously exchanged and ciphering 3DES with the shared 3DES key, the PIN connected to the private key used for the digital signature operation (VERIFY PIN command)
- step 8 Ask the SC for the random number that will be used as challenge for the next commands. It is first step for the activation of Secure Messaging (GET CHALLENGE command)
- step 9 Transmit a random number to the SC as a challenge for the next command. It is the second step for the activation of Secure Messaging (GIVE CHALLENGE command)
- step10 Compute and send, using the two random numbers previously exchanged and ciphering 3DES with the shared 3DES key, the data buffer ciphered using the selected private key and receive the result signature (PSO CDS - Perform Security Operation Compute Digital Signature command)

IV. RESULTS

In this section we present the main experimental results of this work. We use the CPS testing environment to show how external commands interleave with the globally legal commands in a SC based digital signature process. The experiments are carried out with two Common Criteria certified SCs from STM-Incrypto34 and Infineon-CardOs [20], [21]. The main results are shown in figs. 6, 7, and 8 and tables I - IV. The left (right) column of fig. 3 presents the 10 (5) steps of the digital signatures processes with the Incrypto (Infineon) SCs (see section III for the details). Note that we do not count the initial "RESET" and have given similar steps in both processes the same label, although the APDUs associated with these steps may be different.

$P_2$ represents the digital signature process associated with the Infineon SC and the globally legal APDUs of each step are given in table I (here RN is short for "random number"). Fig. 6 and table II show how the commands "Get Challenge" and "Give Challenge" from steps (1,5) and (1,6) from process $P_1$ interleave with steps (2,1) to (2,5) of process $P_2$ on the Infineon SC (central column in fig. 3). No error message is generated, and the process $P_2$ terminates as if no interference occurred. In fact, our experiments show that "Get Challenge" and "Give Challenge" commands of process $P_1$ can interleave with process $P_2$ before and after all of its steps. In this case, the interleaving commands of two globally legal digital signature processes create a result whose trustworthiness has not been assured. Because a user cannot distinguish this situation from one in which no anomaly occurred, this problem might undermine the overall trustworthiness of SC use in an interoperable environment. Furthermore, sending the "Get Challenge" and "Give Challenge" commands repeatedly to the SC could be used by an attacker to put a digital signature process effectively on hold.

TABLE I
GLOBALLY LEGAL APDUS OF PROCESS $P_2$ IN FIG. 3 (HEXADECIMAL REPRESENTATION)

| Node | Commands | Globally legal APDUs | | | | | |
|---|---|---|---|---|---|---|---|
| | | cla | ins | p1 | p2 | lc | data | le |
| 2,1 | Master File | 00 | A4 | 00 | 00 | 00 | - | FF |
| 2,2 | MSE Restore | 00 | 22 | F3 | 30 | 00 | - | 00 |
| 2,3 | MSE Set | 00 | 22 | F1 | B6 | 05 | 4D 00 83 01 31 | 00 |
| 2,4 | Verify | 0C | 20 | 00 | 90 | 04 | PIN | 00 |
| 2,5 | PSO_CDS | 0C | 2A | 9E | 9A | 75 | 00-74 | FF |

TABLE II
GLOBALLY LEGAL AND MODIFIED APDUS OF PROCESS $P_2$ IN FIG. 3 (HEXADECIMAL REPRESENTATION)

| Node | Commands | Globally legal and **modified** APDUs | | | | | |
|---|---|---|---|---|---|---|---|
| | | cla | ins | p1 | p2 | lc | data | le |
| 2,1 | Master File | 00 | A4 | 00 | 00 | 00 | - | FF |
| 2,2 | MSE Restore | 00 | 22 | F3 | 30 | 00 | - | 00 |
| 2,3 | MSE Set | 00 | 22 | F1 | B6 | 05 | 4D 00 83 01 31 | 00 |
| **1,5** | **Get Challenge** | **00** | **84** | **00** | **00** | **00** | **-** | **08** |
| **1,6** | **Give Challenge** | **80** | **86** | **00** | **00** | **08** | **RN** | **00** |
| 2,4 | Verify | 0C | 20 | 00 | 90 | 04 | PIN | 00 |
| 2,5 | PSO_CDS | 0C | 2A | 9E | 9A | 75 | 00-74 | FF |

The results shown in fig. 4 were obtained with the Incrypto SC. As above, $P_1$ represents the digital signature process associated with this SC and the globally legal APDUs of each step are given in table III. Process $P_2$ contains APDUs that are either slightly or substantially different from the globally legal APDUs in $P_1$ (see table IV, the modified parts are printed in bold font). In particular, certain APDUs are not documented for the Incrypto SC: these APDUs are therefore labeled as "undefined". The exact sequence of APDUs in $P_2$ is not part of a single digital signature process on any SC we are aware of. Nevertheless, these commands could well be part of such processes implemented on one or several different SCs.

The command sequence executed in our experiments is shown in the central column of fig. 7. Although this executed process contains six additional commands (five of them "undefined") and four modified commands, it terminates without any error message. In addition, the sequence can be looped back to the first node (1,1) "Master File" after any step of the executed process and afterwards continue until the end. These examples show how drastically digital signature processes can be modified via interleaving commands without the associated anomalies being recognized. An interoperable environment that does not address this issue may not be considered trustworthy and may have vulnerabilities that potential attackers could seek to exploit.

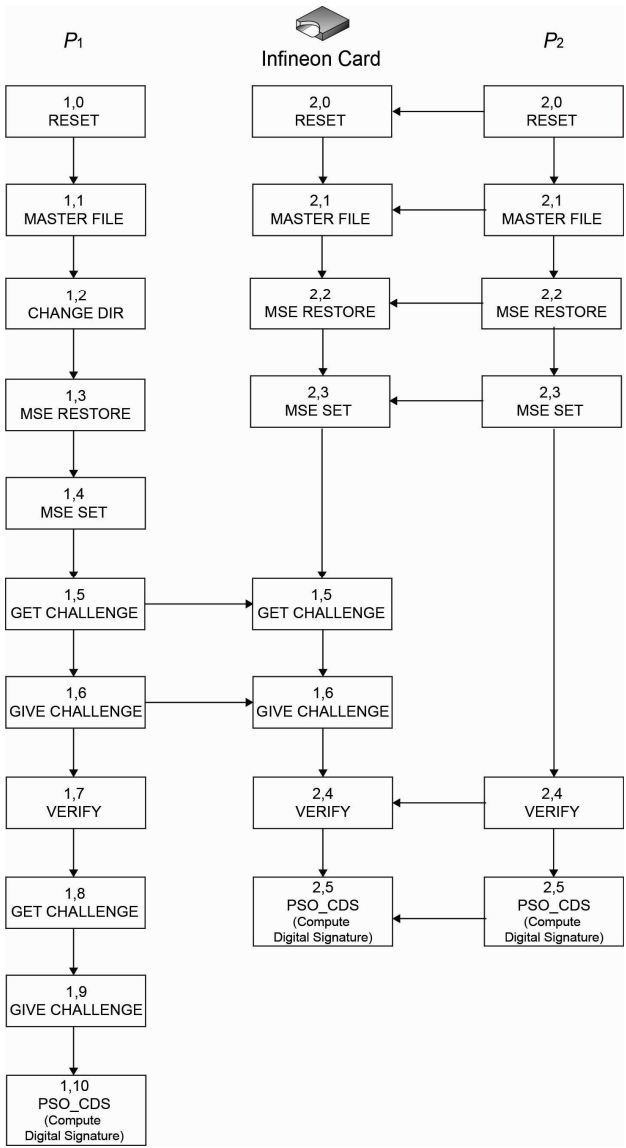

Fig. 6 Two Concurrent Signing Session, Infineon Smartcard

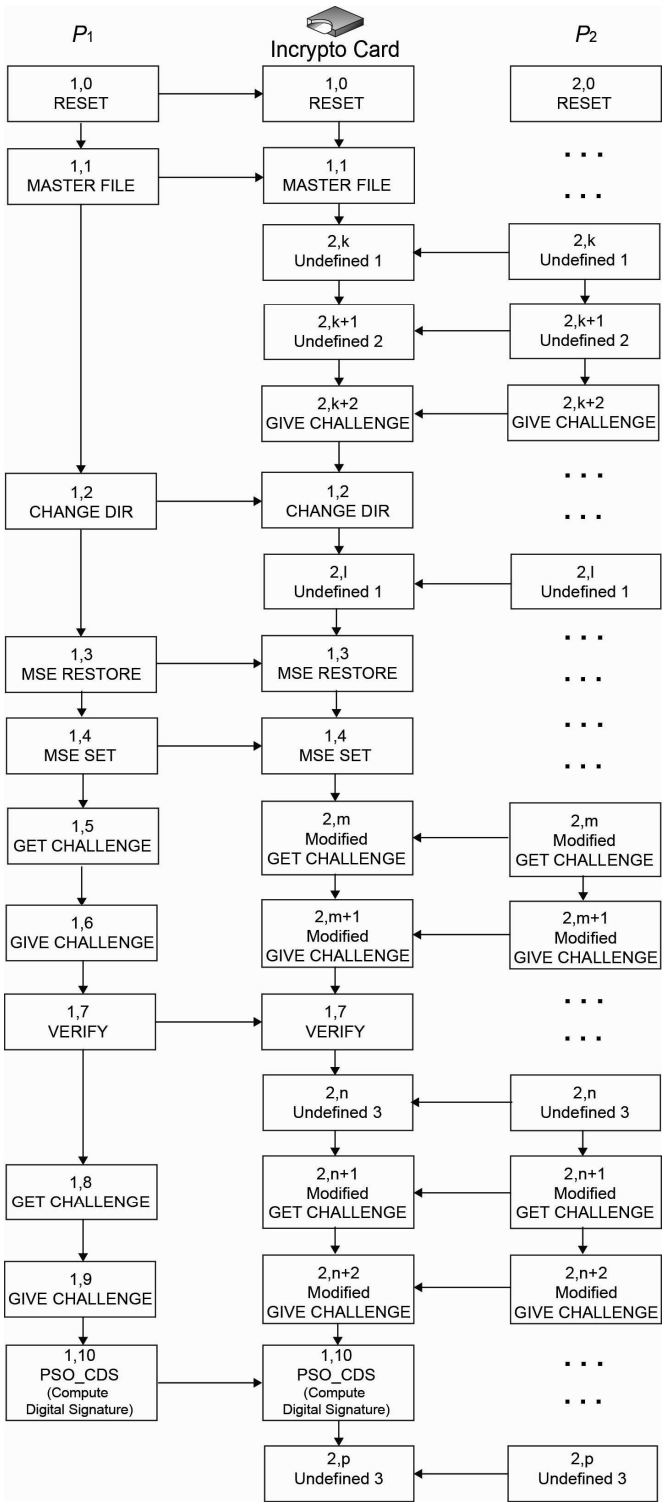

Fig. 7 Two Concurrent Signing Session, Incrypto Smartcard

TABLE III
GLOBALLY LEGAL APDUS OF PROCESS $P_1$ IN FIG. 4 (HEXADECIMAL REPRESENTATION)

| Node | Commands | Globally legal APDUs | | | | | | |
|---|---|---|---|---|---|---|---|---|
| | | cla | ins | p1 | p2 | lc | data | le |
| 1,1 | Master File | 00 | A4 | 00 | 00 | 00 | - | FF |
| 1,2 | Change Dir | 00 | A4 | 00 | 00 | 02 | 14 00 | FF |
| 1,3 | MSE Restore | 00 | 22 | F3 | 03 | 00 | - | 00 |
| 1,4 | MSE Set | 00 | 22 | F1 | B6 | 03 | 83 01 10 | 00 |
| 1,5 | Get Challenge | 00 | 84 | 00 | 00 | 00 | - | 08 |
| 1,6 | Give Challenge | 80 | 86 | 00 | 00 | 08 | RN | 00 |
| 1,7 | Verify | 0C | 20 | 00 | 9A | 08 | PIN | 00 |
| 1,8 | Get Challenge | 00 | 84 | 00 | 00 | 00 | - | 08 |
| 1,9 | Give Challenge | 80 | 86 | 00 | 00 | 08 | RN | 00 |
| 1,10 | PSO_CDS | 0C | 2A | 9E | 9A | 75 | 00-74 | FF |

TABLE IV
GLOBALLY LEGAL AND MODIFIED APDUS OF THE EXECUTED PROCESS OF FIG. 4 (HEXADECIMAL REPRESENTATION)

| Node | Commands | Globally legal and **modified** APDUs | | | | | | |
|---|---|---|---|---|---|---|---|---|
| | | cla | ins | p1 | p2 | lc | data | le |
| 1,1 | Master File | 00 | A4 | 00 | 00 | 00 | - | FF |
| **2,k** | **Undefined1** | **81** | **86** | **00** | **00** | **02** | **14 00** | **00** |
| **2,k+1** | **Undefined2** | **8F** | **86** | **00** | **00** | **02** | **14 00** | **00** |
| **2,k+2** | **Give Challenge** | **80** | **86** | **AC** | **45** | **08** | **RN** | **00** |
| 1,2 | Change Dir | 00 | A4 | 00 | 00 | 02 | 14 00 | FF |
| **2,l** | **Undefined1** | **81** | **86** | **00** | **00** | **02** | **14 00** | **00** |
| 1,3 | MSE Restore | 00 | 22 | F3 | 03 | 00 | - | 00 |
| 1,4 | MSE Set | 00 | 22 | F1 | B6 | 03 | 83 01 10 | 00 |
| **2,m** | **Get Challenge** | 00 | 84 | **BD** | **17** | 00 | - | 08 |
| **2,m+1** | **Give Challenge** | 80 | 86 | **AC** | **45** | 08 | RN | 00 |
| 1,7 | Verify | 0C | 20 | 00 | 9A | 08 | PIN | 00 |
| **2,n** | **Undefined3** | **8C** | **86** | **00** | **00** | **02** | **14 00** | **00** |
| **2,n+1** | **Get Challenge** | 00 | 84 | **BD** | **17** | 00 | - | 08 |
| **2,n+2** | **Give Challenge** | 80 | 86 | **AC** | **45** | 08 | RN | 00 |
| 1,10 | PSO_CDS | 0C | 2A | 9E | 9A | 75 | 00-74 | FF |
| **2,p** | **Undefined3** | **8C** | **86** | **00** | **00** | **02** | **14 00** | **00** |

Finally, we would like to point out a problem caused by interleaving commands that has considerable consequences even though an error is generated. In this experiment (shown in fig. 8 and table V), $P_2$ contains the "MSE Erase" command. This command is usually not part of a digital signature process as it erases the Security Environment Object (SEO); however, it is conceivable that this command is used by an application interacting with the middleware for some purpose. It may then accidentally, or even in an attack, interleave with a digital signature process like $P_1$. We observe experimentally that "MSE Erase", executed as shown in the central column of fig. 8, erases the SEO on the SC without warning, and the digital signature process generates an error after the next step $S_{1,3}$. The digital signature function of the SC is herewith permanently destroyed and a physical replacement of the SC is required. In principal, such vulnerability could be systematically exploited in an attack on all SCs issued by the digital signature service provider because neither PIN nor PUK is required to execute the "MSE Erase" command.

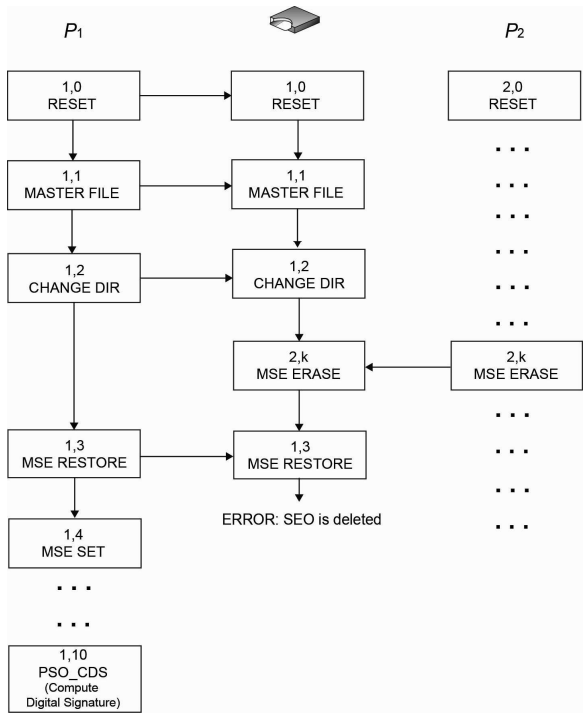

Fig. 8 Signing session with interleaving "MSE Erase" command

TABLE V
MODIFIED APDUs OF PROCESS $P_2$ IN FIG. 5 (HEXADECIMAL REPRESENTATION)

| Node | Commands | Modified APDUs | | | | | | |
|---|---|---|---|---|---|---|---|---|
| | | cla | ins | p1 | p2 | lc | data | le |
| 1,1 | Master File | 00 | A4 | 00 | 00 | 00 | - | FF |
| 1,2 | Change Dir | 00 | A4 | 00 | 00 | 02 | 14 00 | FF |
| **2,k** | **MSE Erase** | **00** | **22** | **F4** | **03** | **00** | **-** | **00** |
| 1,3 | MSE Restore | 00 | 22 | F3 | 03 | 00 | - | 00 |
| MSE Restore returns ERROR, process is terminated ||||||||

## V. CONCLUSION

The experiments described above show that the problem of interleaving command sequences is serious and that it must be addressed to ensure a secure and trustworthy environment for SC interoperability.

As stated in the introduction, in previous work [17], [18] a C-Murphi model checker [22] has been used to test SC behavior in the presence of disturbed commands. Model checking can address extended systems which can assume millions of different states [23] and can in principal be used to identify anomalies. However, the complexity of the verification increases exponentially if interleaving commands are to be taken into account: assume that for every step of two digital signature processes, the input command has only a 16 bits and assume that the two signature processes consist of 10 steps each. Even under this strong simplification, a brute force model checker may be required to make more than $\binom{20}{10} * 2^{16+16}$ tests. This is due to the fact that in this approach all possible sequences that can be obtained by mixing the two signature processes are generated. Note that in an interoperable environment, possibly tens, if not hundreds, of applications may interact concurrently with various SC types via some middleware. As a result, a brute force model checking approach is clearly not a viable solution, especially if it is operated on real SCs as illustrated in the experiments described above where the execution of a single command can take up to 1 second.

In future research, we plan to extend the model checking approach to avoid brute force testing and to identify errors and anomalies effectively. This can be done if one prevents the model checker from searching through all possible sequences of anomalies and errors by taking the results of the already existing CC certification into account. Such an efficient model checker can then be integrated into a middleware as a "watch-dog" to identify an anomaly as it occurs and to prevent computational chains with two or more anomalies. In this case, it will be possible to extend the Common Criteria to certify the anomaly-free interoperability of several SC applications interacting via a middleware with different SC types. conclusion section is not required. Although a conclusion may review the main points of the paper, do not replicate the abstract as the conclusion. A conclusion might elaborate on the importance of the work or suggest applications and extensions.

REFERENCES

[1] N. G. Olve, V. Vimarlund, M. Agerbo, "Evaluation as multi-actor trade-off – a challenge in introducing ICT innovations in the health sector", in *Proceedings of the 4th WSEAS International Conference on E-Activities*, Miami, Florida, USA. WSEAS 2005, pp. 31-48.
[2] C. Popescu, A. Mitu, D. Uta, "Impact measurement for Civitas Success project", in *Proceedings of Recent Researches in Urban Sustainability and Green Development, the 2nd International Conference on Urban Sustainability, Cultural Sustainability, Green Development, Green Structures and Clean Cars (USCUDAR '11)*, Prague, Czech Republic. WSEAS, 2011, pp. 166-171.
[3] C. Toma, M. Popa, C. Boja, "Smart Card based Solution for Non-Repudiation in GSM WAP Applications", in WSEAS Transactions on Computers, issues 5, vol.7. WSEAS 2008, pp. 453-462.
[4] M. Talamo, M. Galinium, C. H. Schunck, F. Arcieri, "Interleaving Command Sequences: a Thread to Secure Smartcard Interoperability", in *Proceedings of the 10th International Conference on Information Security and Privacy (ISP'11)*, Jakarta, Indonesia. WSEAS 2011, pp. 102-107.

[5] M. Talamo, M. Galinium, C. H. Schunck, F. Arcieri, "Integrating Secure Messaging into OpenSC", *in the 2nd International Conference on Computer and Management (CAMAN 2012)*, IEEE, to be published.

[6] *Common Criteria for Information Technology Security Evaluation, version 3.1*, Common Criteria Std. CCMB-2009-07-001, CCMB-2009-07-002, CCMB-2009-07-003, Rev. 3 FINAL, 2009.

[7] *Secure signature-creation devices "EAL 4+"*, European Committee for Standardization (CEN) Std. CWA 14169, 2004.

[8] *Identification cards – Integrated circuit cards Part 4: Organization, security and commands for interchange,*, International Organization for Standardization Std. ISO/IEC 7816-4:2005, Jan. 2005.

[9] *Identification cards – Integrated circuit cards programming interfaces – Part 3: Application programming interface*, International Organization for Standardization Std. ISO/IEC 24727-3:2008, Dec. 2008.

[10] T. Schwarzhoff, et al, "Government smart card interoperability specification, version 2.1", United Stated of America National Institute for Standards and Technology (NIST), Tech. Rep. 6887, 2003.

[11] A. Tauber, B. Zwattendorfer, T. Zefferer, Y. Mazhari, and E. Chamakiotis, "Toward interoperability: An architecture for pan-European eid-based authentication services", in *EGOVIS 2010*, 2010, pp. 120-133.

[12] A. Kazerooni, M. Adlband, O. Mahdiyar, "Application of Public Key Infrastructure in E-Business", in *Proceedings of Recent Researches in Applied Informatics & Remote Sensing, 7th WSEAS International Conference on Applied Computer Science*, Penang, Malaysia. WSEAS 2011, pp. 189-193

[13] M. Y. Siyal, "A Biometric Based E-Security System for Internet-based Applications", in *Proceedings of the 2002 WSEAS International Conference on Electronics, Control & Signal Processing and WSEAS International Conference on e-activities*, Singapore. WSEAS 2002.

[14] T. Cucinotta, M. Di Natale, and D. Corcoran, "A protocol for programmable smart cards", in *Proceedings of the 14th International Workshop on Database and Expert Systems Applications (DEXA03)*. IEEE Computer Society, 2003.

[15] S. Lee, Y, Kim, J. Cho, K. Jung, "An Interoperable Payment Protocol for the Public Transit Fare Payment System", in *Proceedings of the 2002 WSEAS International Conference on Information Security, Hardware/Software Codesign, E-Commerce and Computer Networks*, Rio de Janeiro, Brazil. WSEAS, 2002, pp. 1441-1445.

[16] D. Hüehnlein and M. Bach, "How to use iso/iec 24727-3 with arbitrary smart cards", in *TrustBus 2007*, ser. LNCS vol. 4657. Springer Verlag, 2007, pp. 280-289.

[17] M. Talamo, et al, "Robustness and interoperability problems in security devices", in *Proceedings of 4th International Conferences on Information Security and Cryptology (INSCRYPT) 2008*, 2008.

[18] M. Talamo, et al, "Verifying extended criteria for interoperability of security devices", in *Proceedings of 3rd International Symposium on Information Security, IS08*, ser. LNCS vol. 5332. Springer, 2008, pp. 1131-1139.

[19] W. Rankl and W. Effing, *Smart Card Handbook*, 4th ed. West Sussex, UK: Wiley, 2010.

[20] Siemens, "Certification report cardos v4.2 cns with application for digital signature."

[21] *Secure Signature Creation Device Incrypto34v2 from ST INCARD S.r.l*, Bundesamt für Sicherheit in der Informationtechnik Std. BSI-DSZ-CC-0202-2005, 2005.

[22] G. Della Penna. (2005, Aug.) The CMurphi Verifier. [Online]. Available: http://www.di.univaq.it/gdellape/murphi/cmurphi.php.

[23] E. M. Clarke, O. Grumberg, and D. A. Peled, *Model Checking*. The MIT Press, 1999.